\documentclass{article}
\usepackage{graphicx}
\begin{document}

\title{On the Structure of the Effective Potential
for a Spherical Wormhole}
\author{N. Montelongo Garc\'ia$^{1}$ and T. Zannias$^{2}$\\
nadiezhda@ifm.umich.mx$^{1}$,~zannias@ifm.umich.mx$^{2}$\\
Instituto de F\'isica y Matem\'aticas\\
Universidad Michoacana de San Nicol\'as de Hidalgo\\A.P. 2-82, 58040
 Morelia, Mich.\\ MEXICO}
\maketitle


\begin{abstract}

\noindent The structure of the effective potential $V$ describing
causal geodesics near the throat of an arbitrary spherical wormhole
is analyzed. Einstein's equations relative to a set of regular
coordinates covering a vicinity of the throat imply that any
spherical wormhole can be constructed from solutions of an effective
initial value problem with the throat serving as an initial value
surface. The initial data involve matter variables, the area $A(0)$
of the throat and  the gradient  $\Lambda(0)$ of the red shift
factor on the throat. Whenever $\Lambda(0)=0$, the effective
potential $V$ has a critical point on the throat. Conditions upon
the data are derived ensuring that the critical point is a local
minimum (resp. maximum). For particular families of
Quasi-Schwarzschild wormholes, $V$ exhibits a local minimum on the
throat independently upon the energy $E$ and angular momentum $L^2 $
of the test particles and thus such wormholes admit stable circular
timelike and null geodesics on the throat. For families of Chaplygin
wormholes, we show that such geodesics are unstable. Based on a
suitable power series representation of the metric, properties of
$V$ away from the throat are obtained that are useful for the
analysis of accretion disks and radiation processes near the throat
of any spherical wormhole.
\end{abstract}

\section{Introduction  \label{I}}

\noindent Wormholes, like black holes are topologically non trivial
solutions of Einstein's equations. The former do not possess an
event horizon and this property combined with their non trivial
topology endows them with remarkable properties. If a wormhole is
considered as a handle connecting two distant regions of the same
universe, it offers the opportunity for interstellar travel
\cite{Morris-Thorne-a} and it can act as a time machine
\cite{Morris-Thorne-b}. If it connects two ``different universes''
transfer of matter and radiation from one ``universe'' to the other
is a possibility. This property lead the authors of \cite{Kardashev}
to put forward a radical hypothesis: active galactic nuclei and
other compact objects may be entrances to wormhole throats. Whether
this hypothesis reflects reality, eventually will be decided by
observations combined with analysis of astrophysical processes near
the wormhole throats. In a recent work \cite{Damour-Solodukhin} (see
also \cite{NSolodukhin}) some intriguing properties of wormholes
have been pointed out. Many features of black holes can be closely
mimicked by a family of Quasi-Schwarzschild wormholes \cite{FW}.
Those features include quasinormal mode ringing
\cite{Kokkotas-Schmidt}, the well known $337$ Ohms finite surface
resistivity of an event horizon \cite{Damour} and even some aspects
of Hawking radiation \cite{Hawking}.

In the works of \cite{Kardashev} and \cite{Damour-Solodukhin}, the
behavior of causal geodesics in the vicinity of the throat plays an
important role. In \cite{Kardashev} it has been suggested that the
existence of geodesics oscillating through the throat, may lead to
the observational identification of a wormhole throat. The family of
wormholes employed in \cite{Damour-Solodukhin}, exhibits two worth
noticing properties: They admit stable circular timelike and null
geodesics on the throat and furthermore allow geodesics to oscillate
with respect to the throat. These properties should  be contrasted
with the behavior of geodesics on the spherical wormholes
\cite{K-essence} admitted by massless K-essence \cite{K-essence}.
Spherical wormholes of this theory fall into two families: the so
called zero mass family and a second family having the property that
the two asymptotically flat ends possess ADM masses of opposite
sign. For the later family, the effective potential exhibits no
extrema on the throat and its structure shows absence of geodesics
oscillating through the throat
 \cite{K-essence3}. These contradictory conclusions
that one arrives at from the behavior of geodesics on the family of
Quasi-Schwarzschild wormholes considered in \cite{Damour-Solodukhin}
and the behavior of geodesics on K-essence wormholes, call for an
understanding of the factors influencing the potential near the
throat of an arbitrary spherical wormhole.

The present paper analyzes such issue. In the next section, we
demonstrate that spherical wormholes can be constructed from
solutions of an effective initial value problem (IVP here after)
with the throat serving as an initial value surface. In sec.
(\ref{III}) we introduce the effective potential $V$ and formulate
criteria guaranteeing the presence or absence of stable geodesics on
the throat while in sec. (\ref{IV}) we discuss the structure of $V$
for Quasi-Schwarzschild wormholes. A series representation of the
metric valid in a vicinity containing the throat is derived in sec.
(\ref{V}). This representation offers insights regarding the
structure of $V$ away from the throat. We finish the paper with a
brief summary of the results and open problems.

\section{Constructing Spherical wormholes\label{II}}

\noindent We start with a Gaussian chart covering an open vicinity
of the throat so that $\textbf{g}$ can be written as:

\begin{eqnarray}
\textbf{g}=-e^{2\Phi(l)}dt^{2}+dl^2+r^{2}(l)d\Omega^2, l\,\,\in\,\,
(-\alpha,\alpha).\label{p}
\end{eqnarray}

The throat is located at $l=0$  so that $r(l=0)\neq 0$,
$\frac{dr(l)}{dl}|_{0}=0,$ $\frac{d^2r(l)}{dl^2}|_{0}>0$ and
$\Phi(l)$ is assumed to be smooth \cite{Note1}. Relative to this
chart, we consider a stress tensor $T_{\alpha\beta}$ that decomposes
according to:

\begin{eqnarray}
\nonumber T_{\alpha\beta}&=&\rho(l)
c^2u_{\alpha}u_{\beta}-\tau(l)X_{\alpha}X_{\beta}
\\
&+& P(l)(Y_{\alpha}Y_{\beta}+Z_{\alpha}Z_{\beta}),\label{A1}\\
\nonumber &&\textbf{u}=e^{-\Phi(l)}\frac{\partial}{\partial t},
X=\frac{\partial}{\partial l},\,\, Y=\frac{1}{r(l)}\frac{\partial}
{\partial\theta},\\
\nonumber &&Z=\frac{1}{r(l)\sin\theta}
\frac{\partial}{\partial\phi},
\end{eqnarray}

\nonumber where $(\rho(l) c^2,\tau(l), P(l))$ stand for the energy
density, tension and tangential pressure  as measured by the Killing
observers. In the gauge of (\ref{p}), Einstein's eqs
$G_{\alpha\beta}=\hat{k}T_{\alpha\beta}$ and
$\nabla_{\alpha}T^{\alpha\beta}=0$  can be cast in the form:

\begin{eqnarray}
\frac{dr(l)}{dl}&=&\frac{1}{2}K(l)r(l),\label{w}\\
\frac{d K(l)}{dl}&=&-\frac{3}{4}K^2(l)+\frac{1}{r^2}
-\hat{k}\rho(l)c^2,\label{x}\\
\frac{d\Lambda(l)}{dl}&=&-\frac{1}{2}K(l)\Lambda(l)
+\frac{1}{2}\hat{k}\rho(l)c^2-
\frac{1}{2r^2(l)}\\
\nonumber &+&\frac{K^2(l)}{8} -\Lambda^2(l)
+\hat{k}P(l),\label{y}\\
\frac{d\tau(l)}{dl}&=&[\rho(l)c^2-\tau(l)] \Lambda(l) -[\tau(l)
\\
\nonumber
&+&P(l)]K(l),\label{z}\\
-\frac{1}{r^2(l)}&+&\frac{K^2(l)}{4}+\hat{k}\tau(l)
+\Lambda(l)K(l)=0,\label{nueva}
\end{eqnarray}

\noindent where $K(l)=\frac{2}{r(l)}\frac{dr(l)}{dl}$ is the trace
of the extrinsic curvature of the $SO(3)$ orbits as embedded in
$t=$const hypersurfaces and $\Lambda(l)=\frac{d\Phi(l)}{dl}$. As a
consequence of (\ref{w}-\ref{z}), it follows that (\ref{nueva}) is a
constraint and if it is satisfied at an $l_0$, then it is satisfied
for all $l$. Evaluating (\ref{nueva}) on the throat we obtain
$\hat{k}\tau(0)=r^{-2}(0)$, while from (\ref{x}) it follows that
$A(l)=4\pi r^{2}(l)$ exhibits a local minimum at $l=0$ provided
$\hat{k}r^2(0)\rho(0)c^2<1$ or equivalently
$(\tau(0)-\rho(0)c^2)>0.$ Thus any solution of (\ref{w}-\ref{z})
subject to:

\begin{eqnarray}
r(0)=\left(\frac{A(0)}{4\pi}\right)^{\frac{1}{2}},K(0)=0,
\Lambda(0),\tau(0)=\frac{1}{\hat{k}r^{2}(0)},\label{X}
\end{eqnarray}

\noindent with $\rho(0)$ chosen so that $(\tau(0)-\rho(0)c^2)>0$ and
$\Lambda(0)$ arbitrary, describes a local wormhole. Notice that the
constraint and dynamical eqs (\ref{w}-\ref{z}) do not restrict
$\Lambda(0)$ and thus together with $A(0)$ may be viewed as the
degrees of freedom characterizing the throat.
\newline
\newline
In the absence of equations of state or dependence of $(\rho c^2,$
$\tau,$ $ P)$ upon fundamental fields \cite{Note2}, the system
(\ref{w}-\ref{z}) simplifies. Two of the unknown functions can be
a-priori specified \cite{Note3} and on grounds of mathematical
simplicity we shall specify $(\rho c^{2}, \Lambda)$, referred here
after as the free data. For that case, eq. (\ref{y}) written in the
form:

\begin{eqnarray}
\nonumber \hat{k}
P(l)&=&\frac{d\Lambda(l)}{dl}+\Lambda^{2}(l)+\frac{1}{2}
\frac{dK(l)}{dl} +\frac{K^{2}(l)}{4}
\\
&+&\frac{K(l)\Lambda(l)}{2},\label{c8}
\end{eqnarray}

\noindent implies that $P(l)$ is determined once  $(r(l),K(l))$ are
known. As long as $\rho c^{2}$ is continuous and $\Lambda$ is
continuous differentiable on $(-\alpha,\alpha)$ then
Picard-Lindellof's theorem \cite{PHartman} applied to
(\ref{w},\ref{x},\ref{z}) guarantees the existence of a unique
$C^{1}$ solution defined on $[-b,b]\subset(-\alpha,\alpha)$.  In
\cite{MZ} it was shown that under mild restrictions upon $(\rho
c^{2},\Lambda)$, local solutions of  (\ref{w},\ref{x},\ref{z}) can
be extended so that they represent smooth, asymptotically flat
wormholes \cite{Note4}.

The treatment of (\ref{w}-\ref{z}) has to proceed along different
lines whenever $T_{\alpha\,\beta}$ described by (\ref{A1}) is
attributed to a fluid or defined by an underlying field theory such
as $K$-essence \cite{Note2}. For instance, in the case that
$T_{\alpha\,\beta}$ describes a perfect fluid with four velocity
$\textbf{u}$ parallel to the Killing field then $(\rho c^2,\tau, P)$
are linked by equations of state. For a monoparametric family of
equations of state: $\tau=F(\rho), P=G(\rho)$, eqs
$(\ref{y},\ref{z})$ take the form:

\begin{eqnarray}
\nonumber &&\frac{d\Lambda(l)}{dl}=-\frac{1}{2}K(l)\Lambda(l)
+\frac{1}{2}\hat{k}\rho(l)c^2-
\frac{1}{2r^2(l)}+\frac{K^2(l)}{8}\\
&&-\Lambda^2(l) +\hat{k}G(\rho(l)),\label{n8}
\end{eqnarray}
\begin{eqnarray}
\nonumber \frac{d
F(\rho)}{d\rho}\frac{d\rho(l)}{dl}&=&[\rho(l)c^{2}-F(\rho(l))]\Lambda(l)
\\&-&[F(\rho(l))+G(\rho(l))]K(l).\label{n7}
\end{eqnarray}

\noindent Again  as long as $F,G$ are smooth functions,
Picard-Lindeloff's theorem applied to (\ref{w},\ref{x},
\ref{n8},\ref{n7}) combined with ({\ref{X}}) assures the existence
of a local solution. However properties of the maximal solution are
by no means obvious. Depending upon the equations of state,
(\ref{n7}) can be a highly non linear equation  and the possibility
that  while the data are propagating off the throat a singularity in
the solution will develop is not excluded \cite{Note5}. For the case
where $T_{\alpha\beta}$ is associated with fields, the situation is
not getting any easier. Eq. (\ref{z}) which determines the tension,
will involve second derivatives of the fields and similar comments
regarding  the maximal solution apply in this case as well
\cite{Note6}.

Despite the absence of a global wormhole solution of
(\ref{w}-\ref{z}), by appealing to the local solutions predicted by
the Picard-Lindeloff we can analyze the structure of the effective
potential describing geodesics in a vicinity of the throat.

\section{Structure of the effective potential on the throat\label{III}}

\noindent On the background of (\ref{p}), any timelike or null
geodesic $x^{\mu}(\mu)=(t(\mu),l(\mu),\theta(\mu),\phi(\mu))$
satisfies:

\begin{eqnarray}
&&\frac{d t(\mu)}{d\mu}=-\frac{E}{g_{00}},\,\,
\frac{d\phi(\mu)}{d\mu}=\frac{L}{g_{\phi\phi}},\,\,\theta(\mu)
=\frac{\pi}{2},\\
&&\left(\frac{dl(\mu)}{d\mu}\right)^{2}=\frac{1}{e^{2\Phi(l)}}
[E^{2}-V(l,L^{2})],
\end{eqnarray}

\noindent where $\mu$ denotes proper time for a timelike and an
affine parameter for a null geodesic. $(E,L)$ are constants along a
particular geodesic and we have made use of spherical symmetry to
set: $\theta(\mu)=\frac{\pi}{2}$. The function $V(l,L^2)$ is
referred as the effective potential and is defined by:

\begin{eqnarray}
V(l,L^2)=e^{2\Phi(l)}(k+\frac{4\pi L^2}{A(l)}),\label{A2}
\end{eqnarray}

\noindent where $A(l)=4\pi r^2(l)$ and $k=1$ ( $k=0$ ) for timelike
(null) geodesics. It follows from (\ref{A2}) that:

\begin{eqnarray}
\nonumber &&\frac{\partial V(l,L^{2})}{\partial
l}=e^{2\Phi(l)}\left[2\Lambda(l) (k+\frac{4\pi L^2}{A(l)})\right.
\\
&&\left.-\frac{4\pi L^2}{A^2(l)}\frac{d A(l)}{dl}\right],\label{ZZZ}
\end{eqnarray}

\noindent and thus if $\Lambda(0)\neq 0$ then $V(l,L^2)$ has a
regular value on the throat for all $L^2\neq 0$ \cite{Note7}.
Geodesics can pass from one asymptotic region to the other or they
may have a turning point on the throat. Moreover depending on the
maxima and minima of $V(l,L^2)$ away from the throat, there may
exist geodesics
 oscillating through the throat \cite{Note8}.

However for $\Lambda(0)=0$,  $V(l,L^2)$ has a critical point on the
throat regardless of the value of $L^2$ and whether $k=1$ or $k=0$.
 Evaluating the second derivative
of $V(l,L^2)$ on the throat, we obtain:

\begin{eqnarray}
&&\frac{\partial^2 V(l,L^{2})}{\partial
l^2}\mid_{0}=e^{2\Phi(0)}\left[2\frac{d\Lambda(l)}{dl}(k+\frac{4\pi
L^2}{A(l)})\right.
\\
\nonumber && \left. -\frac{4\pi
L^2}{A^2(l)}\frac{d^2A(l)}{dl^2}\right]\mid_{0}.\label{m1}
\end{eqnarray}

\noindent For the case where $(\rho c^{2},\Lambda)$ are free data,
it reduces to:

\begin{eqnarray}
&&\frac{\partial^{2} V(l,L^{2})}{\partial
l^{2}}|_{0}=e^{2\Phi(0)}\left[
2\frac{d\Lambda(l)}{dl}\left(k+\frac{4\pi L^{2}}{A(l)}\right)\right.
\\
\nonumber &&\left.-\frac{4\pi L^{2}}{A(l)}\left(\frac{1}{r^{2}(l)}-
\hat{k}\rho(l)c^{2}\right)\right]|_{0},\label{n3}
\end{eqnarray}

and thus if $\frac{d\Lambda(l)}{dl}|_{0}\leq 0$ then $V(l,L^{2})$
possesses a local maximum for all $L^{2}$. In this case any timelike
or null geodesic on the throat would be unstable. On the other hand,
whenever $\frac{d\Lambda(l)}{dl}|_{0}>0$, the nature of the extremum
depends upon $L^{2}$ and the form of $\frac{d\Lambda(l)}{dl}|_{0}$.

For wormholes supported by a perfect fluid, the right hand side of
(\ref{m1}) takes the form:

\begin{eqnarray}
\nonumber
 &&\frac{\partial^2 V(l,L^{2})}{\partial
l^2}\mid_{0}=e^{2\Phi(0)}\hat{k}[k(2P(0)-a(0))
\\
&+&\frac{8\pi L^2}{A(0)}(P(0) -a(0))],\label{noche}
\end{eqnarray}

\noindent where $a(0)\equiv \tau(0)-\rho(0)c^{2}$. Therefore,
whenever the data satisfy $P(0)\geq a(0)\equiv \tau(0)-\rho(0)c^2$,
the extremum of $V(l,L^2)$  is a local minimum regardless of the
value of $L^{2}$ and whether $k=1$ or $k=0$. For other values of
$P(0)$ the nature of the extremum can be easily inferred from
(\ref{noche}).

In the next section we shall consider geodesic motion for the family
of Quasi-Schwarzschild wormholes \cite{MVisser}.

\section{The effective potential for Quasi-Schwarzschild
wormholes\label{IV}}

\noindent A Quasi-Schwarzschild wormhole is a spherical wormhole
with the property that the intrinsic geometry of the hypersurfaces
orthogonal to the timelike Killing field is isometric to the
corresponding geometry of the positive mass Schwarzschild manifold
\cite{Note9}. If $\gamma$ stands for the induced metric on the
$t=$cons hypersurfaces and $R(\gamma)$ its scalar curvature, the
hamiltonian constraint for Einstein's equations imply that
$R(\gamma)=2\hat{k}\rho c^2$ and thus for any Quasi-Schwarzschild
wormhole $\rho\equiv 0$. This property
 simplifies the system
(\ref{w}-\ref{nueva}). In fact if we define $r(l)$ via:

\begin{eqnarray}
 \frac{dr(l)}{dl}=\pm\sqrt{1-\frac{r_{0}}{r(l)}},\, r(0)=r_{0},\,\,l\,\in\,(-\infty,\infty),\label{dia}
\end{eqnarray}

\noindent it can be verified that  this $r(l)$ combined with

\begin{eqnarray}
\nonumber &&K(l)=\frac{2}{r(l)}\frac{dr(l)}{dl}=\pm\frac{2}{r(l)}
\left[1-\frac{r_{0}}{r(l)}\right]^{1/2},
\\
&&\,\,\,l\,\,\,\in\,\,\,(-\infty,\infty), \label{n1}
\end{eqnarray}

\noindent satisfies (\ref{w},\ref{x}) and the required initial
conditions ( we adopt the notation that the $(+)$ sign corresponds
to $l\,\,\,\in\,\, (0,\infty)$ and $(-)$ for $l\,\,\, \in\,\,
(-\infty,0)$). Combining these expressions for $(r(l),K(l))$ with
any $C^{1}$ function $\Lambda(l)$ subject to:
$\lim_{l\to\pm\infty}\,\,\Lambda(l)=O(l^{-2})$ \cite{Note11}, eqs
(\ref{y},\ref{nueva}) are satisfied provided:

\begin{eqnarray}
\nonumber
&&\hat{k}\tau(l)=\frac{r_{0}}{r^{3}(l)}-\frac{2\Lambda(l)}{r(l)}
\left[1-\frac{r_{0}}{r(l)}\right]^{1/2},\\
&& l \in [0,\infty),\label{m4}
\\
\nonumber &&\hat{k}P(l)=\frac{d\Lambda(l)}{dl}+\Lambda^{2}(l)
+\frac{\Lambda(l)}{r(l)}\left[1-\frac{r_{0}}{r(l)}\right]^{1/2}+
\\
&&\frac{r_{0}}{2r^{3}(l)},\,\,\, l\,\,\, \in
\,\,\,[0,\infty),\label{m5}
\end{eqnarray}

\noindent while the corresponding $\hat{k}\tau(l), \hat{k} P(l)$ for
$l\,\,\, \in\,\,\, (-\infty,0]$ are obtained from (\ref{m4},
\ref{m5})
 by changing the sign in front of the
linear term in $\Lambda$. Finally the metric has the form:

\begin{eqnarray}
\nonumber
&&\textbf{g}=-e^{2\Phi(0)+2\int_{0}^{l}\Lambda(l')dl'}dt^{2}+dl^{2}
+r^{2}(l)d\Omega^{2},\\
&& l\,\in\,(-\infty,\infty),\label{m2}
\end{eqnarray}

\noindent with $\Phi(0)$ an arbitrary constant. If on the sheet with
$l\,\,\in\,\, (0,\infty)$ resp. $l\,\,\,\in\,\,\, (-\infty,0)$ we
eliminate the $l$ coordinate in favor of $r$ then the spatial metric
takes the form:

$$ \gamma=\frac{dr^{2}}{1-\frac{r_{0}}{r}}+r^{2}d\Omega^{2},\,
\,\,r\,\,\in\,\,\,(r_{0},\infty),$$

\noindent showing that the ends $l\,\,\,\to\,\,\infty$ resp.
$l\,\,\to -\infty$ , are asymptotically flat ends with equal ADM
masses. Orbiting test particles in the asymptotic regions are
coupled to the so called Komar mass (often referred in the
literature as the Schwarzschild mass) which is determined by the
asymptotic behavior of $\Lambda(l)$.

For any Quasi-Schwarzschild wormhole, it follows from (\ref{ZZZ})
that the critical points  $l_e$ of $V(l,L^2)$ satisfy:

\begin{eqnarray}
\Lambda(l)(k r^{2}(l)+L^{2})=\pm
\frac{L^{2}}{r(l)}\left(1-\frac{r_{0}}{r(l)}\right)^{1/2},\label{NN}
\end{eqnarray}

\noindent and using (\ref{m1}) the nature of the critical point on
the throat can be determined. It is worth to notice that due to the
freedom in the choice of $\Lambda(l)$, one can give $V(l,L^2)$ any
desirable structure
 and bellow we shall discuss a few examples.

Setting $\Lambda(l)=0$ in ( \ref{m4},\ref {m5},\ref{m2}) we recover
the Bronnikov-Kim family of wormholes (derived in \cite{Bronnikov1}
via different techniques and motivations). For this family, it
follows from (\ref{A2}) or from (\ref{NN}) that $V(l,L^2)$ has a
unique global maximum on the throat for all $L^2$ and thus circular
timelike of null geodesics on the throat are unstable. The
Damour-Solodukhin family \cite{Damour-Solodukhin}, mentioned in the
introduction section, is generated by:

\begin{eqnarray}
\Lambda(l)&=&\frac{r_{0}}{2r^{2}(l)}\frac{\sqrt{f}}{(f+\lambda^{2})},
\,\,f(l)=1-\frac{r_{0}}{r(l)},\\
\nonumber && \lambda\neq 0,\,\,\ l\,\, \in\,\, [0,\infty),\label{m3}
\end{eqnarray}

\noindent and $\Lambda(l)=-\Lambda(-l)$. For this $\Lambda(l)$, we
obtain from (\ref{m4},\ref{m5}):

\begin{eqnarray}
\hat{k} P(l)&=&\frac{r_{0}^{2}\lambda^{2}}{4r^{4}(l)}
\frac{1}{(1-\frac{r_{0}}{r(l)}+\lambda^{2})^{2}} \label{n4}
\\
\nonumber &+& \frac{r_{0}\lambda^{2}}{2r^{3}(l)}
\frac{1}{(1-\frac{r_{0}}{r(l)}+\lambda^{2})},\,\,
 l\,\, \in\,\, (-\infty,\infty),\\
\hat{k}\tau(l)&=&\frac{r_{0}\lambda^{2}}{r^{3}(l)}
\frac{1}{(1-\frac{r_{0}}{r(l)}+\lambda^{2})},\,\,\, l \,\,\in\,\,
(-\infty,\infty),
\end{eqnarray}

\noindent whereas (\ref{n3}) yields:

\begin{eqnarray}
\nonumber &&\frac{\partial^{2}V(l,L^{2})}{\partial
l^{2}}|_{0}=e^{2\Phi(0)}\left[\frac{4\pi
L^{2}}{A(0)r^{2}(0)}\left(\frac{1}{2\lambda^{2}}-1\right)\right.
\\
&+&\left.\frac{k}{2r_{0}^{2}\lambda^{2}}\right].
\end{eqnarray}

\noindent Thus, as long as $\lambda^{2}<\frac{1}{2}$ this family
admits stable circular timelike and null geodesics on the throat in
agreement with the conclusions of ref.\cite{Damour-Solodukhin}. Away
from the throat the effects of non vanishing  $\lambda^{2}$ can be
easily worked out.

\section{On the effective potential near the throat\label{V}}

\noindent In an effort to get insights in the properties of
$V(l,L^2)$ away from the throat, in this section we construct a
power series solution of $(\ref{w}-\ref{nueva})$ valid in a vicinity
containing the throat. For simplicity we shall treat only  wormholes
with reflective symmetry \cite{Note4}, although the following
analysis can be extended to wormholes lacking this symmetry. We
shall treat first the case where $(\rho c^2,\Lambda)$ are smooth
free data so that \cite{Note10}:

\begin{eqnarray}
\nonumber \rho(l)c^2&=&\rho(0)c^2+\rho(1)l
+\rho(2)l^2+\rho(3)l^3\\&&+O(l^4),\label{nua}\\
\nonumber \Lambda(l)&=&\Lambda(0)+\Lambda(1)l+ \Lambda(2)l^2
+\Lambda(3)l^3\\
&+& O(l^{5}).\label{nub}
\end{eqnarray}

\noindent According to the results of section (\ref{II}), we choose
$\Lambda(0)=0, \hat{k}\tau(0)=r^{-2}(0)$ so that
$\tau(0)>\rho(0)c^2$ and set:
$\rho(1)=\rho(3)=\Lambda(0)=\Lambda(2)=0$. It is a matter of algebra
to show that a power series solution of (\ref{w}-\ref{z}) has the
form:

\begin{eqnarray}
&&r(l)=r(0) +\frac{\hat{k}r(0)}{4}(\tau(0)-\rho(0)c^2)l^2+O(l^4),
\label{AAAAAA}\\
&&K(l)=\hat{k}(\tau(0)-\rho(0)c^2)l+K(3)l^3+O(l^5),\label{n10}
\end{eqnarray}

\begin{eqnarray}
\hat{k}P(l)&=&[\Lambda(1)+\frac{\hat{k}}{2}(\tau(0)-\rho(0)c^{2})]
\\
\nonumber &+&
\left[\Lambda^{2}(1)+2\Lambda(3)-\frac{\hat{k}\rho(2)}{2}\right.
+\frac{\hat{k}}{2} (\tau(0)-\rho(0)c^{2})
\\
\nonumber &&(\Lambda(1)-\frac{\hat{k}\tau(0)}{2})
-\left.\frac{\hat{k}^{2}}{8}(\tau(0)-\rho(0)c^{2})^{2}\right]l^2+O(l^3),
\end{eqnarray}

\begin{eqnarray}
\nonumber \hat{k}\tau(l)&=&\hat{k}\tau(0)
-\hat{k}[\tau(0)-\rho(0)c^2][\hat{k}P(0)
\\
&+&\frac{\hat{k}\tau(0)+\hat{k}\rho(0)c^2}{4}]l^2+O(l^4),
\label{BBBBBB}
\end{eqnarray}

\noindent with $K(3)$ described by:

\begin{eqnarray}
K(3)&=&-\frac{\hat{k}^2}{12}[(\tau(0)
-\rho(0)c^2)(5\tau(0)-3\rho(0)c^2)]
\\
\nonumber &+&\frac{\hat{k}\rho(2)}{3}.
\end{eqnarray}

This series implies that the metric $\textbf{g}$ can be written in
the form:

\begin{eqnarray}
&&\textbf{g}=-[1+\Lambda(1)l^{2}
+\frac{\Lambda(3)}{2}l^{4}+O(l^{6})]dt^2+dl^2\\
\nonumber &&+
 r^{2}(0)[1+\frac{\hat{k}}{4}(\tau(0)
 -\rho(0)c^2)l^{2}+O(l^4)]d\Omega^{2},
\end{eqnarray}

\noindent where in above we have set $\Phi(0)=0$. From (\ref{ZZZ}),
the critical points of $V(l,L^2)$ are identified as the roots of the
equation:

\begin{eqnarray}
-L^2 K(l)+2\Lambda(l) (k r^2(l)+L^2)=0,\label{EEEE}
\end{eqnarray}

\noindent and in terms of the series solution it yields a lengthly
algebraic cubic equation. One root occurs at $l=0$ and the other two
are either real and of opposite sign or they
 are imaginary. We have checked that there exist data so
that on the non vanishing real roots, $V(l,L^2)$ exhibits local
maxima and moreover exist geodesics that oscillate through the
throat, a conclusion that verifies the claim in \cite{Kardashev}.

Finally we shall briefly consider Chaplygin wormholes. Such
wormholes are supported by a perfect fluid stress tensor
$T_{\alpha\,\beta} $ so that the isotropic pressure $P$ and energy
density ${\rho} c^2$ obey: $P(\rho)=-\frac{A}{c^{2}\rho},\,\,\,A\neq
0$ \cite{Note12,Eiora1}. Setting $P=-\tau$, the constraint
$\tau(0)-c^{2}\rho(0)>0$ is satisfied provide $A>\rho^{2}(0)$. Again
we assume:

\begin{eqnarray}
c^{2}\rho(l)=c^{2}\rho(0)+\rho(2)l^{2}+O(l^{4}),\label{n6}
\end{eqnarray}

\noindent and thus

\begin{eqnarray}
\tau(l)=\frac{A}{c^{2}\rho(0)}\left[1-\frac{\rho(2)}{c^{2}\rho(0)}l^{2}+
O(l^{4})\right].
\end{eqnarray}

Substituting these expressions into (\ref{w}-\ref{x}) we obtain:

\begin{eqnarray}
\nonumber
r(l)&=&r(0)+\frac{1}{4r(0)}[1-\hat{k}\rho(0)c^{2}r^{2}(0)]l^{2}\\
&+&O(l^{4}),\label{n9}\\
K(l)&=&\frac{4l}{r(0)}\left[\frac{1}{4r(0)}[1-\hat{k}\rho(0)c^{2}r^
{2}(0)]+O(l^{2})\right],\label{nn1}\\
\nonumber
\Lambda(l)&=&-\left[\frac{1}{2r^{2}(0)}[1-\hat{k}c^{2}\rho(0)r^{2}(0)]
+\frac{\hat{k}A}{c^{2}\rho(0)}\right]l
\\
&+&O(l^{2}),\label{nn2}
\end{eqnarray}

\noindent resulting into:

\begin{eqnarray}
\nonumber
&&\textbf{g}=-\left[1-\left(\frac{1-\hat{k}c^{2}\rho(0)r^{2}(0)}{2r^{2}(0)}
+\frac{\hat{k}A}{c^{2}\rho(0)}\right)l^{2}+O(l^{4})\right]dt^{2}
\\
&&+dl^{2}+r^{2}(0)\left[
1+\frac{1-\hat{k}\rho(0)c^{2}r^{2}(0)}{2r^{2}(0)}l^{2}
+O(l^{4})\right]d\Omega^{2}.
\end{eqnarray}

\noindent For this solution we find:

\begin{eqnarray}
&&\frac{\partial^{2}V(l,L^{2})}{\partial l^{2}}|=
-e^{2\phi(0)}\left[ \left(k+\frac{4\pi
L^{2}}{A(0)}\right)\frac{2\hat{k}A}{c^{2}\rho(0)}\right.
\\
&+&\left. \frac{1-\hat{k}c^{2}\rho(0)r^{2}(0)}{r^{2}(0)}\left(k+
\frac{8\pi L^{2}}{A(0)}\right)\right],
\end{eqnarray}

\noindent and since $A$ is positive and
${1-\hat{k}c^{2}\rho(0)r^{2}(0)}>0$, all circular timelike or photon
orbits on the throat are unstable \cite{Note13}. The behavior of
$V(l,L^2)$ away from the throat can be obtained from (\ref{EEEE})
adapted to the series (\ref{n6}-\ref{nn2}).

\section{Discussion\label{VI}}

\noindent In this work a connection between initial data determining
a wormhole and properties of the effective potential $V$ near the
throat is established. Our analysis shows that with suitable
selection of the data, $V$ can acquire any desirable structure as
illustrated by the Quasi-Schwarzschild and Chaplygin wormholes. In
the course of the paper, we have also analyzed the structure of
Quasi-Schwarzschild wormholes. Such wormholes have simple geometries
and this can be useful. For instance, they can be used as
backgrounds where the effects of the  wormhole topology upon
accretion flows and emergent spectra can be addressed. Of course the
stability  of Quasi-Schwarzschild wormholes has to be analyzed and
we hope to discuss some of these issues in the near future.

\section{Acknowledgments\label{VII}}

\noindent This work is partially supported by grants: 4.7-CIC-UMSNH,
$E9507$-COECYT and N. M. G. acknowledges a doctoral fellowship from
CONACYT-Mexico. We would like to express our thanks to Drs.
J.Estevez-Delgado, U.Nucamendi, F.Guzman, J.A.Gonzalez and O.Sarbach
for many stimulating discussions. Particular thanks are due to
F.Guzman and Jose A.Gonzalez and J.Estevez Delgado for reading
carefully the manuscript and for their constructive comments.


\end{document}